\documentclass[12pt,dvips]{article}
\let\section=\subsection     \let\subsection=\subsubsection                
\usepackage{graphicx}

\usepackage[a4paper=true,colorlinks=false,bookmarks=true,%
        pdftitle={Short-ranged radial and tensor correlations in nuclear many-body systems},%
        pdfsubject={Hirschegg 2003 - Nuclear Structure and Dynamics at the Limits},%
        pdfauthor={Thomas Neff <t.neff@gsi.de>}]{hyperref}

\begin{document}


\newcommand{\half}{{\textstyle\frac{1}{2}}}
\newcommand{\quarter}{{\textstyle\frac{1}{4}}}
\newcommand{\const}{\mathit{const}}

\newcommand{\fm}{\mathrm{fm}}
\newcommand{\MeV}{\mathrm{MeV}}
\newcommand{\GeV}{\mathrm{GeV}}


\newcommand{\vecsigma}{\boldsymbol{\sigma}}
\newcommand{\vectau}{\boldsymbol{\tau}}
\newcommand{\vecrho}{\boldsymbol{\rho}}
\newcommand{\vecpi}{\boldsymbol{\pi}}

\newcommand{\eqdef}{\equiv}
\newcommand{\sheq}{\!=\!}

\newcommand{\shcdot}{\!\cdot\!}

\newcommand{\eqdot}{\; .}
\newcommand{\eqcomma}{\; ,}

\renewcommand{\Re}{\operatorname{Re}}
\renewcommand{\Im}{\operatorname{Im}}
\newcommand{\erf}{\operatorname{erf}}
\newcommand{\erfi}{\operatorname{erfi}}

\newcommand{\bra}[1]{\big< \,{#1}\, \big| }

\newcommand{\ket}[1]{\big| \,{#1}\, \big> }

\newcommand{\braket}[2]{\big< \,{#1}\, \big| \,{#2}\, \big> }
\newcommand{\braketa}[2]{\braket{#1}{#2}{}_a}
\newcommand{\braketaa}[2]{{}_a \braket{#1}{#2}{}_a}

\newcommand{\expect}[1]{\big< \, {#1} \, \big>}

\newcommand{\matrixe}[3]{\big< \,{#1}\, \big| \,{#2}\, 
\big| \,{#3}\, \big> }
\newcommand{\matrixea}[3]{\matrixe{#1}{#2}{#3}{}_a}
\newcommand{\matrixeaa}[3]{{}_a\matrixe{#1}{#2}{#3}{}_a}

\newcommand{\comm}[2]{\bigl[ {#1}, {#2} \bigr]_{-} }
\newcommand{\acomm}[2]{\bigl[ {#1}, {#2} \bigr]_{+} }
\newcommand{\commpm}[2]{\bigl[ {#1}, {#2} \bigr]_{\pm} }
\newcommand{\commmp}[2]{\bigl[ {#1}, {#2} \bigr]_{\mp} }

\newcommand{\biggcomm}[2]{\biggl[ {#1}, {#2} \biggr]_{-} }
\newcommand{\biggacomm}[2]{\biggl[ {#1}, {#2} \biggr]_{+} }
\newcommand{\biggcommpm}[2]{\biggl[ {#1}, {#2} \biggr]_{\pm} }
\newcommand{\biggcommmp}[2]{\biggl[ {#1}, {#2} \biggr]_{\mp} }

\newcommand{\totd}[2]{\ensuremath{ \frac{d {#1}} {d {#2}} }}

\newcommand{\totdd}[2]{\ensuremath{ \frac{d^2 {#1}}{d {#2}^2} }}

\newcommand{\partd}[2]{\ensuremath{ \frac{\partial #1}
{\partial #2} }}

\newcommand{\partdd}[2]{\ensuremath{ \frac{\partial^2 #1}
{\partial #2^2} }}

\newcommand{\ddt}[1]{\totd{{#1}}{t}}

\newcommand{\pddt}[1]{\partd{{#1}}{t}}

\newcommand{\dddtt}[1]{\totdd{{#1}}{t}}

\newcommand{\pdddtt}[1]{\partdd{{#1}}{t}}

\newcommand{\intd}[1]{\int\!\mathrm{d}{#1}\;}
\newcommand{\intdt}[1]{\int\!\mathrm{d}^3{#1}\;}

\newcommand{\Lagrangian}{\mathscr{L}}
\newcommand{\Hamiltonian}{\mathscr{H}}

\newcommand{\conj}[1]{{{#1}}^{\star}}
\newcommand{\hermit}[1]{{{#1}}^{\dag}}

\newcommand{\op}[1]{%
    \fontdimen12\textfont3=2pt\fontdimen12\scriptfont3=1.4pt%
    \!\null\mathop{\vphantom{#1}\smash{#1}}\limits_{\!\sim}\null\!}
\newcommand{\hop}[1]{\hermit{\op{#1}}}

\newcommand{\conop}[1]{\op{#1}^{\dagger}}
\newcommand{\desop}[1]{\op{#1}^{\phantom{\dagger}}}

\newcommand{\opid}{\op{1}}

\newcommand{\VecThree}[3]{%
   \left( \begin{matrix}{#1} \\ {#2} \\ {#3} \end{matrix} \right)}

\newcommand{\Trace}{\mathrm{Tr}}

\newcommand{\grad}[1]{\partd{}{#1}}
\newcommand{\gradx}{\partd{}{\vec{x}}}

\newcommand{\rhoone}{\rho^{(1)}}
\newcommand{\rhotwo}{\rho^{(2)}}
\newcommand{\oprhoone}{\op{\rho}^{(1)}}
\newcommand{\oprhotwo}{\op{\rho}^{(2)}}


\newcommand{\STBops}{STB operators}

\newcommand{\eqrrep}{\overset{\scriptstyle{\vec{r}}}{\Rightarrow}}

\newcommand{\eqtwoapp}{\overset{\scriptstyle{[2]}}{=}}

\newcommand{\couple}[3]{\left\{ {#1} \: {#2} \right\}^{(#3)}}
\newcommand{\coupletensor}[3]{\left\{ {#1} \otimes {#2} \right\}^{(#3)}}

\newcommand{\couplev}[3]{({#1}\,{#2})^{(#3)}}

\newcommand{\sixj}[6]{\left\{ \begin{matrix} {#1} & {#3} & {#5}\\ {#2}
& {#4} & {#6} \end{matrix} \right\}}
\newcommand{\ninej}[9]{\left\{ \begin{matrix} {#1} & {#4} & {#7}\\ {#2}
& {#5} & {#8} \\ {#3} & {#6} & {#9} \end{matrix} \right\}}
\newcommand{\cg}[6]{\mathrm{C}\Biggl(\,\begin{matrix} {#1} & {#3} \\ {#2}
& {#4} \end{matrix}\, \Biggr|\Biggl.\, \begin{matrix} {#5} \\
{#6} \end{matrix} \,\Biggr)}

\newcommand{\talmi}[4]{\left< \begin{matrix} {#1} \\ {#2} \end{matrix}
    \right|\left. \begin{matrix} {#3} \\ {#4} \end{matrix} \right>}

\newcommand{\corr}[1]{\hat{#1}}

\newcommand{\cop}[1]{\op{\corr{#1}}}

\newcommand{\opC}{\op{C}}
\newcommand{\hopC}{\hop{C}}
\newcommand{\opCr}{\op{C}_{r}^{\phantom{\dagger}}}
\newcommand{\hopCr}{\op{C}_{r}^{\dagger}}
\newcommand{\opCom}{\op{C}_{\Omega}^{\phantom{\dagger}}}
\newcommand{\hopCom}{\op{C}_{\Omega}^{\dagger}}
\newcommand{\opG}{\op{G}}

\newcommand{\sopLom}{\mathsf{L}_{\Omega}}

\newcommand{\opc}{\op{c}}
\newcommand{\hopc}{\hop{c}}
\newcommand{\opcr}{\op{c}_{r}^{\phantom{\dagger}}}
\newcommand{\hopcr}{\op{c}_{r}^{\dagger}}
\newcommand{\opcom}{\op{c}_{\Omega}^{\phantom{\dagger}}}
\newcommand{\hopcom}{\op{c}_{\Omega}^{\dagger}}
\newcommand{\opg}{\op{g}}
\newcommand{\opgr}{\op{g}_{r}}
\newcommand{\opgom}{\op{g}_{\Omega}}

\newcommand{\opone}[1]{\op{#1}^{[1]}}
\newcommand{\copone}[1]{\cop{#1}^{[1]}}
\newcommand{\optwo}[1]{\op{#1}^{[2]}}
\newcommand{\coptwo}[1]{\cop{#1}^{[2]}}

\newcommand{\icm}{\mathrm{cm}}
\newcommand{\irel}{\mathrm{rel}}

\newcommand{\vecr}{\vec{r}}
\newcommand{\opvecr}{\op{\vec{r}}}
\newcommand{\vecx}{\vec{x}}
\newcommand{\opvecx}{\op{\vec{x}}}
\newcommand{\vecX}{\vec{X}}
\newcommand{\opvecX}{\op{\vec{X}}}
\newcommand{\vecp}{\vec{p}}
\newcommand{\opvecp}{\op{\vec{p}}}
\newcommand{\vecP}{\vec{P}}
\newcommand{\opvecP}{\op{\vec{P}}}
\newcommand{\vecl}{\vec{l}}
\newcommand{\opvecl}{\op{\vec{l}}}

\newcommand{\vecpr}{\vec{p}_{r}}
\newcommand{\opvecpr}{\op{\vec{p}}_{r}}
\newcommand{\pr}{p_{r}}
\newcommand{\oppr}{\op{p}_{r}}

\newcommand{\pom}{p_{\Omega}}
\newcommand{\vecpom}{\vec{p}_{\Omega}}
\newcommand{\opvecpom}{\op{\vec{p}}_{\Omega}}

\newcommand{\lsq}{\vecl^2}
\newcommand{\oplsq}{\op{\vecl}^2}

\newcommand{\gom}{g_{\Omega}}
\newcommand{\srpom}{s_{\!12}(\vecr,\vecpom)}
\newcommand{\opsrpom}{\op{s}_{\!12}(\vecr,\vecpom)}
\newcommand{\opSrpom}{\op{S}_{\!12}(\vecr,\vecpom)}

\newcommand{\Rp}{R_{+}}
\newcommand{\Rm}{R_{-}}
\newcommand{\Rpm}{R_{\pm}}
\newcommand{\Rmp}{R_{\mp}}
\newcommand{\metricRp}{\mathcal{R}_{+}}
\newcommand{\metricRm}{\mathcal{R}_{-}}
\newcommand{\metricRpm}{\mathcal{R}_{\pm}}
\newcommand{\metricRmp}{\mathcal{R}_{\mp}}

\newcommand{\cmur}{\hat{\mu}_{r}}
\newcommand{\cmuom}{\hat{\mu}_{\Omega}}
\newcommand{\cu}{\hat{u}}

\newcommand{\opSone}{\op{S}^{(1)}}
\newcommand{\opStwo}{\op{S}^{(2)}}

\newcommand{\Pinot}{\Pi_{0}}
\newcommand{\Pione}{\Pi_{1}}
\newcommand{\ls}{\vecl\shcdot\vec{s}}
\newcommand{\lssq}{(\ls)^2}
\newcommand{\srr}{s_{\!12}(\unitvec{r},\unitvec{r})}
\newcommand{\sll}{s_{\!12}(\vecl,\vecl)}
\newcommand{\spompom}{s_{\!12}(\vecpom,\vecpom)}
\newcommand{\sbarpompom}{\bar{s}_{\!12}(\vecpom,\vecpom)}

\newcommand{\opPinot}{\op{\Pi}_{0}}
\newcommand{\opPione}{\op{\Pi}_{1}}
\newcommand{\opls}{\op{\vecl}\shcdot\op{\vec{s}}}
\newcommand{\oplssq}{(\opls)^2}
\newcommand{\opsrr}{\op{s}_{\!12}(\unitvec{r},\unitvec{r})}
\newcommand{\opsll}{\op{s}_{\!12}(\vecl,\vecl)}
\newcommand{\opspompom}{\op{s}_{\!12}(\vecpom,\vecpom)}
\newcommand{\opsbarpompom}{\op{\bar{s}}_{\!12}(\vecpom,\vecpom)}

\newcommand{\opLsq}{\op{\vec{L}}^2}

\newcommand{\lone}{\op{l}^{(1)}}
\newcommand{\rrtwo}{\couplev{\op{\hat{r}}}{\op{\hat{r}}}{2}}
\newcommand{\rpomtwo}{\couplev{\op{r}}{\op{\pom}}{2}}
\newcommand{\lltwo}{\couplev{\op{l}}{\op{l}}{2}}
\newcommand{\pompomtwo}{\couplev{\op{\pom}}{\op{\pom}}{2}}
\newcommand{\pompombartwo}{(\overline{\op{\pom}\,\op{\pom}})^{(2)}}


\newcommand{\crhoone}{\corr{\rho}^{(1)}}
\newcommand{\crhotwo}{\corr{\rho}^{(2)}}

\newcommand{\coprhoone}{\op{\corr{\rho}}^{(1)}}
\newcommand{\coprhotwo}{\op{\corr{\rho}}^{(2)}}


\newcommand{\FMD}{Fermionic Molecular Dynamics}


\newcommand{\chemical}[2]{\ensuremath{{}^{#1}\mathrm{#2}}}

\newcommand{\Hefour}{\chemical{4}{He}}
\newcommand{\Osixteen}{\chemical{16}{O}}
\newcommand{\Cafourty}{\chemical{40}{Ca}}

\newcommand{\calC}{\mathscr{C}}
\newcommand{\calB}{\mathscr{B}}
\newcommand{\calH}{\mathscr{H}}
\newcommand{\calO}{\mathsf{o}}


\newcommand{\mat}[1]{\mathbf{\mathsf{#1}}}

\newcommand{\cone}[1]{\corr{#1}^{[1]}}
\newcommand{\ctwo}[1]{\corr{#1}^{[2]}}


\newcommand{\opS}{\op{S}}
\newcommand{\opA}{\op{\mathcal{A}}}

\newcommand{\opH}{\op{H}}
\newcommand{\copH}{\cop{H}}
\newcommand{\copHone}{\cop{H}^{[1]}}
\newcommand{\copHtwo}{\cop{H}^{[2]}}

\newcommand{\opT}{\op{T}}
\newcommand{\copT}{\cop{T}}
\newcommand{\copTone}{\cop{T}^{[1]}}
\newcommand{\copTtwo}{\cop{T}^{[2]}}

\newcommand{\copTtworad}{\copTtwo_{r}}
\newcommand{\copTtwograd}{\copTtwo_{\vec{r}}}
\newcommand{\copTtwopot}{\copTtwo_{\mathit{pot}}}

\newcommand{\opTintr}{\op{T}_{\mathit{intr}}}
\newcommand{\opHintr}{\op{H}_{\mathit{intr}}}

\newcommand{\opTcm}{\op{T}_{\mathit{cm}}}
\newcommand{\opTcmone}{\op{T}_{\mathit{cm}}^{[1]}}
\newcommand{\opTcmtwo}{\op{T}_{\mathit{cm}}^{[2]}}

\newcommand{\opV}{\op{V}}
\newcommand{\opVC}{\op{V}_C}
\newcommand{\opVT}{\op{V}_T}
\newcommand{\opVLS}{\op{V}_{LS}}

\newcommand{\oprhoonex}{\oprhoone(\vec{x})}
\newcommand{\oprhoonek}{\oprhoone(\vec{k})}

\newcommand{\oprhopone}{\op{\rho}^{(1)}_{\mathit{k}}}
\newcommand{\oprhonone}{\op{\rho}^{(1)}_{\mathit{n}}}
\newcommand{\oprhoponex}{\op{\rho}^{(1)}_{\mathit{p}}(\vec{x})}
\newcommand{\oprhononex}{\op{\rho}^{(1)}_{\mathit{n}}(\vec{x})}
\newcommand{\oprhoponek}{\op{\rho}^{(1)}_{\mathit{p}}(\vec{k})}
\newcommand{\oprhononek}{\op{\rho}^{(1)}_{\mathit{n}}(\vec{k})}

\newcommand{\opx}{\op{\vec{x}}}
\newcommand{\opk}{\op{\vec{k}}}
\newcommand{\opl}{\op{\vec{l}}}

\newcommand{\opK}{\op{\vec{K}}}
\newcommand{\opX}{\op{\vec{X}}}
\newcommand{\opXp}{\opX_p}
\newcommand{\opXn}{\opX_n}

\newcommand{\opPp}{\op{P}^{p}}
\newcommand{\opPn}{\op{P}^{n}}

\newcommand{\Ppkl}{P^p_{kl}}
\newcommand{\Ppkm}{P^p_{km}}
\newcommand{\Ppln}{P^p_{ln}}
\newcommand{\Ppkk}{P^p_{kk}}
\newcommand{\Ppll}{P^p_{ll}}
\newcommand{\Pnkl}{P^n_{kl}}
\newcommand{\Pnkm}{P^n_{km}}
\newcommand{\Pnln}{P^n_{ln}}

\newcommand{\opXsq}{\op{\vec{X}}^2}
\newcommand{\opXpsq}{\op{\vec{X}}_p^2}
\newcommand{\opXnsq}{\op{\vec{X}}_n^2}

\newcommand{\PiQ}{\ensuremath{\op{\Pi}_1^Q}}
\newcommand{\PiQq}{\ensuremath{\op{\Pi}_1^{\bar{Q}}}}


\newcommand{\ketQ}{\ket{Q}}
\newcommand{\braQ}{\bra{Q}}
\newcommand{\ketQt}{\ket{Q(t)}}
\newcommand{\braQt}{\bra{Q(t)}}
\newcommand{\ketQtp}{\ket{Q'(t)}}

\newcommand{\ketcQ}{\ket{\corr{Q}}}
\newcommand{\bracQ}{\bra{\corr{Q}}}
\newcommand{\ketcQt}{\ket{\corr{Q}(t)}}
\newcommand{\bracQt}{\bra{\corr{Q}(t)}}
\newcommand{\ketcQtp}{\ket{\corr{Q}'(t)}}

\newcommand{\ketcQi}{\ket{\corr{Q}^{i}}}
\newcommand{\ketcQj}{\ket{\corr{Q}^{j}}}
\newcommand{\bracQi}{\bra{\corr{Q}^{i}}}
\newcommand{\bracQj}{\bra{\corr{Q}^{j}}}

\newcommand{\braketQ}{\braket{Q}{Q}}
\newcommand{\matrixeQ}[1]{\matrixe{Q}{#1}{Q}}
\newcommand{\braketQt}{\braket{Q(t)}{Q(t)}}
\newcommand{\matrixeQt}[1]{\matrixe{Q(t)}{#1}{Q(t)}}

\newcommand{\braketQp}{\braket{Q}{Q'}}
\newcommand{\matrixeQp}[1]{\matrixe{Q}{#1}{Q'}}
\newcommand{\braketQtp}{\braket{Q(t)}{Q'(t)}}
\newcommand{\matrixeQtp}[1]{\matrixe{Q(t)}{#1}{Q'(t)}}

\newcommand{\braketcQ}{\braket{\corr{Q}}{\corr{Q}}}
\newcommand{\matrixecQ}[1]{\matrixe{\corr{Q}}{#1}{\corr{Q}}}
\newcommand{\braketcQt}{\braket{\corr{Q}(t)}{\corr{Q}(t)}}
\newcommand{\matrixecQt}[1]{\matrixe{\corr{Q}(t)}{#1}{\corr{Q}(t)}}

\newcommand{\braketcQij}{\braket{\corr{Q}^{i}}{\corr{Q}^{j}}}
\newcommand{\matrixecQij}[1]{\matrixe{\corr{Q}^{i}}{#1}{\corr{Q}^{j}}}

\newcommand{\braketcQp}{\braket{\corr{Q}}{\corr{Q}'}}
\newcommand{\matrixecQp}[1]{\matrixe{\corr{Q}}{#1}{\corr{Q}'}}
\newcommand{\braketcQtp}{\braket{\corr{Q}(t)}{\corr{Q}'(t)}}
\newcommand{\matrixecQtp}[1]{\matrixe{\corr{Q}(t)}{#1}{\corr{Q}'(t)}}


\newcommand{\qckj}{\conj{q}_{k,j}}
\newcommand{\qcmj}{\conj{q}_{m,j}}
\newcommand{\ddqckj}{\partd{}{\qckj}}
\newcommand{\ddqcmj}{\partd{}{\qcmj}}

\newcommand{\ketq}{\ket{q(t)}}
\newcommand{\braq}{\bra{q(t)}}

\newcommand{\qkl}{q_k, q_l}
\newcommand{\qlk}{q_l, q_k}
\newcommand{\qmn}{q_m, q_n}
\newcommand{\qkm}{q_k, q_m}
\newcommand{\qmk}{q_m, q_k}
\newcommand{\qml}{q_m, q_l}
\newcommand{\qlm}{q_l, q_m}
\newcommand{\ooa}{\mathsf{o}_{mk}\mathsf{o}_{nl}-\mathsf{o}_{ml}\mathsf{o}_{nk}}

\newcommand{\ab}{a,\vec{b}}

\newcommand{\abt}{a(t), \vec{b}(t)}
\newcommand{\abtk}{a_k(t), \vec{b}_k(t)}

\newcommand{\gammakl}{\gamma_{kl}}
\newcommand{\omegakl}{\omega_{kl}}
\newcommand{\zetak}{\zeta_k}
\newcommand{\zetal}{\zeta_l}
\newcommand{\etakl}{\eta_{kl}}


\newcommand{\spinup}{\chi^{\scriptscriptstyle\uparrow}}
\newcommand{\spinlo}{\chi^{\scriptscriptstyle\downarrow}}
\newcommand{\spinorchi}[1]{\chiup_{#1},\chilo_{#1}}


\newcommand{\ak}{a_k}
\newcommand{\ack}{\conj{a}_k}
\newcommand{\al}{a_l}
\newcommand{\acl}{\conj{a}_l}
\newcommand{\am}{a_m}
\newcommand{\an}{a_n}

\newcommand{\bk}{\vec{b}_k}
\newcommand{\bck}{\conj{\bk}}
\newcommand{\bl}{\vec{b}_l}
\newcommand{\bcl}{\conj{\bl}}
\newcommand{\bm}{\vec{b}_m}
\newcommand{\bn}{\vec{b}_n}

\newcommand{\chik}{\chi_k}
\newcommand{\chil}{\chi_l}
\newcommand{\chim}{\chi_m}
\newcommand{\chin}{\chi_n}

\newcommand{\xik}{\xi_k}
\newcommand{\xil}{\xi_l}
\newcommand{\xim}{\xi_m}
\newcommand{\xin}{\xi_n}

\newcommand{\abk}{\ak \bk}
\newcommand{\abl}{\al \bl}
\newcommand{\abkl}{\ak\bk,\al\bl}
\newcommand{\abmn}{\am \bm, \an \bn}

\newcommand{\abchik}{\ak \bk \chik}
\newcommand{\abchil}{\al \bl \chil}
\newcommand{\abchikl}{\ak \bk \chik, \al \bl \chil}
\newcommand{\abchimn}{\am \bm \chim, \an \bn \chin}

\newcommand{\abchixik}{\ak \bk \chik \xik}
\newcommand{\abchixil}{\al \bl \chil \xil}
\newcommand{\abchixikl}{\ak \bk \chik \xik, \al \bl \chil \xil}
\newcommand{\abchiximn}{\am \bm \chim \xim, \an \bn \chin \xin}

\newcommand{\Rkl}{R_{kl}}
\newcommand{\Rkm}{R_{km}}
\newcommand{\Rln}{R_{ln}}
\newcommand{\Skl}{S_{kl}}
\newcommand{\Skm}{S_{km}}
\newcommand{\Sln}{S_{ln}}
\newcommand{\Tkl}{T_{kl}}
\newcommand{\Tkm}{T_{km}}
\newcommand{\Tln}{T_{ln}}
\newcommand{\Qkl}{Q_{kl}}
\newcommand{\Qkm}{Q_{km}}
\newcommand{\Qln}{Q_{ln}}
\newcommand{\Olk}{\mathsf{o}_{lk}}
\newcommand{\Omk}{\mathsf{o}_{mk}}
\newcommand{\Onl}{\mathsf{o}_{nl}}

\newcommand{\lambdakl}{\lambda_{kl}}
\newcommand{\lambdakm}{\lambda_{km}}
\newcommand{\lambdaln}{\lambda_{ln}}
\newcommand{\lambdaklmn}{\lambda_{klmn}}
\newcommand{\alphakl}{\alpha_{kl}}
\newcommand{\alphakm}{\alpha_{km}}
\newcommand{\alphaln}{\alpha_{ln}}
\newcommand{\alphaklmn}{\alpha_{klmn}}
\newcommand{\alphaklkl}{\alpha_{klkl}}
\newcommand{\pikl}{\vecpi_{kl}}
\newcommand{\pikm}{\vecpi_{km}}
\newcommand{\piln}{\vecpi_{ln}}
\newcommand{\piklmn}{\vecpi_{klmn}}
\newcommand{\rhokl}{\vecrho_{kl}}
\newcommand{\rhokm}{\vecrho_{km}}
\newcommand{\rholn}{\vecrho_{ln}}
\newcommand{\rhoklmn}{\vecrho_{klmn}}
\newcommand{\rhoklkl}{\vecrho_{klkl}}
\newcommand{\arhoklkl}{\bigl| \rhoklkl \bigr|}
\newcommand{\thetaklmn}{\theta_{klmn}}
\newcommand{\betaklmn}{\beta_{klmn}}

\newcommand{\sigkl}{\vec{\sigma}_{kl}}
\newcommand{\sigkm}{\vec{\sigma}_{km}}
\newcommand{\sigln}{\vec{\sigma}_{ln}}
\newcommand{\Sklmn}{\vec{S}_{klmn}}

\newcommand{\akap}{\alphaklmn+\kappa}
\newcommand{\iakap}{\frac{1}{\akap}}
\newcommand{\kapakap}{\frac{\kappa}{\akap}}
\newcommand{\rhosqakap}{\frac{\rhoklmn^2}{2(\alphaklmn+\kappa)}}
\newcommand{\exprhosqakap}{\exp \biggl\{ - \rhosqakap \biggr\}}
\newcommand{\akapi}{\alphaklmn+\kappa_i}
\newcommand{\iakapi}{\frac{1}{\akapi}}
\newcommand{\kapakapi}{\frac{\kappa_i}{\alphaklmn+\kappa_i}}
\newcommand{\rhosqakapi}{\frac{\rhoklmn^2}{2(\alphaklmn+\kappa_i)}}
\newcommand{\exprhosqakapi}{\exp \biggl\{ - \rhosqakapi \biggr\}}

\newcommand{\matrixeabkl}[1]{\matrixe{\abk}{#1}{\abl}}
\newcommand{\matrixeabklmn}[1]{\matrixe{\abkl}{#1}{\abmn}}
\newcommand{\matrixeabchikl}[1]{\matrixe{\abchik}{#1}{\abchil}}
\newcommand{\matrixeabchiklmn}[1]{\matrixe{\abchikl}{#1}{\abchimn}}
\newcommand{\matrixeabchixikl}[1]{\matrixe{\abchixik}{#1}{\abchixil}}
\newcommand{\matrixeabchixiklmn}[1]{\matrixe{\abchixikl}{#1}{\abchiximn}}


\newcommand{\ddack}{\partd{}{\ack}}
\newcommand{\ddal}{\partd{}{\al}}
\newcommand{\ddacl}{\partd{}{\acl}}
\newcommand{\ddam}{\partd{}{\am}}
\newcommand{\ddan}{\partd{}{\an}}

\newcommand{\ddbck}{\partd{}{\bck}}
\newcommand{\ddbl}{\partd{}{\bl}}
\newcommand{\ddbcl}{\partd{}{\bcl}}
\newcommand{\ddbm}{\partd{}{\bm}}
\newcommand{\ddbn}{\partd{}{\bn}}


\newcommand{\optau}{\op{\vectau}}
\newcommand{\opsig}{\op{\vecsigma}}
\newcommand{\tautau}{\vectau\! \cdot\! \vectau}
\newcommand{\sigsig}{\vecsigma\! \cdot \! \vecsigma}
\newcommand{\optautau}{\optau\! \cdot\! \optau}
\newcommand{\opsigsig}{\opsig\! \cdot \! \opsig}
\newcommand{\optensor}{\ensuremath{\op{S}_{12}}}
\newcommand{\opspinorbit}{\ensuremath{\op{\vec{L}}\!\cdot\!\op{\vec{S}}}}


\newcommand{\Xx}{\vec{X},\vec{x}}
\newcommand{\Xpxp}{\vec{X}', \vec{x}'}
\newcommand{\deltaXXpdeltaxxp}{\delta(\vec{X}-\vec{X}') \delta(\vec{x}-\vec{x}')}
\newcommand{\expxsqkapi}{\exp \biggl\{ -\frac{\vec{x}^2}{2 \kappa_i} \biggr\}}

\newcommand{\expabklXx}{\exp \biggl\{ 
        - \frac{\bigl( \vec{X} + \half \vec{x} - \bck \bigr)^2}{2\ack}
        - \frac{\bigl( \vec{X} -\half \vec{x} - \bcl \bigr)^2}{2\acl} 
        \biggr\}}

\newcommand{\expabmnXx}{\exp \biggl\{ 
        - \frac{\bigl( \vec{X} + \half \vec{x} - \bm \bigr)^2}{2\am}
        - \frac{\bigl( \vec{X} -\half \vec{x} - \bn \bigr)^2}{2\an} 
        \biggr\}}


\newcommand{\intdtx}{\intdt{x}}
\newcommand{\intdtX}{\intdt{X}}
\newcommand{\intdtp}{\intdt{p}}
\newcommand{\intdtP}{\intdt{P}}
\newcommand{\intdtk}{\intdt{k}}
\newcommand{\intdtK}{\intdt{K}}

\newcommand{\murad}{\mu_{r}^{\star}}
\newcommand{\mugrad}{\mu}

\newcommand{\Tcspop}{\op{T}_{\mathit{sp}}^{(2)}}
\newcommand{\Tcradop}{\op{T}_{\mathit{rad}}^{(2)}}
\newcommand{\Msp}{M_{\mathit{sp}}}
\newcommand{\Mrad}{M_{\mathit{rad}}}

\newcommand{\gradbkl}{\frac{1}{2} \biggl( \partd{}{\bcl} -
\partd{}{\bck} \biggr)}
\newcommand{\gradbmn}{\frac{1}{2} \biggl( \partd{}{\bn} -
\partd{}{\bm} \biggr)}

\newcommand{\gradabkl}{\biggl( \ack \partd{}{\bck} - \acl \partd{}{\bcl} +
(\bck - \bcl) \biggr)}
\newcommand{\gradabmn}{\biggl( \am \partd{}{\bm} - \an \partd{}{\bn} +
(\bm - \bn) \biggr)}

\newcommand{\Gklmn}{G_{klmn}}
\newcommand{\tGklmn}{\tilde{G}_{klmn}}


\newcommand{\Qcont}{\ensuremath{Q_\mathrm{cont}}}
\newcommand{\ketQcont}{\ensuremath{\ket{\Qcont}}}
\newcommand{\acont}{a_\mathrm{cont}}
\newcommand{\bcont}{\vec{b}_\mathrm{cont}}


\newcommand{\V}{\ensuremath{V}}
\newcommand{\sV}{\ensuremath{V^{\sigma}}}
\newcommand{\tV}{\ensuremath{V^{\tau}}}
\newcommand{\tsV}{\ensuremath{V^{\sigma\tau}}}
\newcommand{\U}{\ensuremath{T_{pot}}}
\newcommand{\sU}{\ensuremath{T_{pot}^{\sigma}}}
\newcommand{\tU}{\ensuremath{T_{pot}^{\tau}}}
\newcommand{\tsU}{\ensuremath{T_{pot}^{\sigma\tau}}}
\newcommand{\VT}{\ensuremath{V_T}}
\newcommand{\VLS}{\ensuremath{V_{LS}}}
\newcommand{\Tgrad}{\ensuremath{T_{\vec{r}}}}
\newcommand{\sTgrad}{\ensuremath{T_{\vec{r}}^{\sigma}}}
\newcommand{\tTgrad}{\ensuremath{T_{\vec{r}}^{\tau}}}
\newcommand{\tsTgrad}{\ensuremath{T_{\vec{r}}^{\sigma\tau}}}
\newcommand{\Trad}{\ensuremath{T_{r}}}
\newcommand{\sTrad}{\ensuremath{T_{r}^{\sigma}}}
\newcommand{\tTrad}{\ensuremath{T_{r}^{\tau}}}
\newcommand{\tsTrad}{\ensuremath{T_{r}^{\sigma\tau}}}

\newcommand{\psiop}{\ensuremath{\op{\Psi}}}
\newcommand{\psiopc}{\ensuremath{\op{\Psi}^{\dagger}}}

\newcommand{\idxij}{i\hspace*{-0.7pt}j}

\begin{center}
   {\large \bf Short-ranged radial and tensor correlations\\
     in nuclear many-body systems}\\[2mm]
     T.~Neff and H.~Feldmeier \\[2mm]
   {\small \it  Gesellschaft f\"ur Schwerionenforschung (GSI) \\
   Postfach 110552, D-64220 Darmstadt, Germany \\[5mm] }
\end{center}

\begin{abstract}\noindent
  The Unitary Correlation Operator Method (UCOM) is applied to
  realistic potentials.  The effects of tensor correlations are
  investigated.  The resulting phase shift equivalent correlated
  interactions are used in the no-core shell model for light nuclei
  and for mean-field calculations in the Fermionic Molecular Dynamics
  model for nuclei up to mass $A=48$.
\end{abstract}

\section{Motivation}

In principle the interaction among hadrons should be derived from
Quantum Chromo Dynamics (QCD), the fundamental theory of the strong
interaction.  However a description of the atomic nucleus in terms of
QCD is still not in sight.  Therefore one starts with so called
realistic nucleon-nucleon (NN) potentials that reproduce phase shifts
up to $E_{lab}=300 \MeV$ and the deuteron properties.  This procedure
is not unique as the NN-interaction is momentum dependent.
Furthermore, exact calculations of the three- and four-nucleon system
show that realistic NN-potentials are not sufficient and genuine
three-body forces are needed to describe the binding energies.

Our aim is to use realistic NN-potentials in nuclear structure
calculations where the many-body Hilbert space is spanned by Slater
determinants.  Although those form in principle a complete basis they
are very badly suited to describe the correlations induced by the
repulsive core and the strong tensor part in the NN-potential
$\op{V}$. A large scale shell model calculation with the bare
interaction for $\chemical{3}{He}$ needs already $50\:\hbar \Omega$
excitations, a Hilbert space of dimension $10^5$, to achieve
convergence.

\section{Unitary Correlation Operator Method (UCOM)}

In order to avoid the large off-diagonal matrix elements that scatter
to high lying shell-model states we propose [1-2] to first perform a
unitary transformation of the Hamiltonian
\begin{equation} \label{eq:corrH}
  \cop{H} = \hop{C} (\op{T}+ \op{V}) \op{C} = 
  \op{T}+\coptwo{H}+\cop{H}^{[3]}+\cdots
\end{equation}
that incorporates the effects of the repulsive and tensor correlations
in the sense of a pre-diagonalization.  In Eq.~(\ref{eq:corrH})
$\op{T}$ denotes the kinetic energy, $\coptwo{H}$ the
two-body, and $\cop{H}^{[3]}$ the three-body part of the
correlated Hamiltonian.  With this unitary transformation the original
eigenvalue problem in terms of the bare Hamiltonian $\op{H}$ and
many-body states $\ket{\corr{\Psi}_n}$ that include short ranged
correlations can be rewritten in terms of a correlated Hamiltonian
$\cop{H}$ and more simple states $\ket{\Psi_n}$ that are Slater
determinants or superpositions of a limited number of those:
\begin{equation}
  \op{H} \ket{\corr{\Psi}_n} = E_n \ket{\corr{\Psi}_n}\quad
  \rightarrow \quad
  \cop{H} \ket{\Psi_n}= \hop{C} \op{H} \op{C} \ket{\Psi_n} = E_n \ket{\Psi_n}
\end{equation}

The correlator $\op{C}=\op{C}_\Omega \op{C}_r$ consists of the unitary
radial correlator
\begin{equation}
  \op{C}_r = \exp \biggl\{ -i \sum_{i<j} \frac{1}{2}
  \Bigl(s(\op{r}_{\idxij})\, \op{p}_{r \idxij} + h.a. \Bigr)\biggr\}
\end{equation}
and the unitary tensor correlator
\begin{equation} \label{eq:opcom}
  \op{C}_\Omega = \exp \biggl\{ -i \sum_{i<j} \frac{3}{2}
   \vartheta(\op{r}_{\idxij})\:\Bigl( (\op{\vec{\sigma}}_i\, \op{\vec{r}}_{\idxij})
  (\op{\vec{\sigma}}_j\, \op{\vec{p}}_{\Omega \idxij}) + h.a.
  \Bigr)\biggr\} \eqdot
\end{equation}

\begin{figure}[tb]
  \begin{center}
    \includegraphics[angle=0,width=0.9\linewidth]{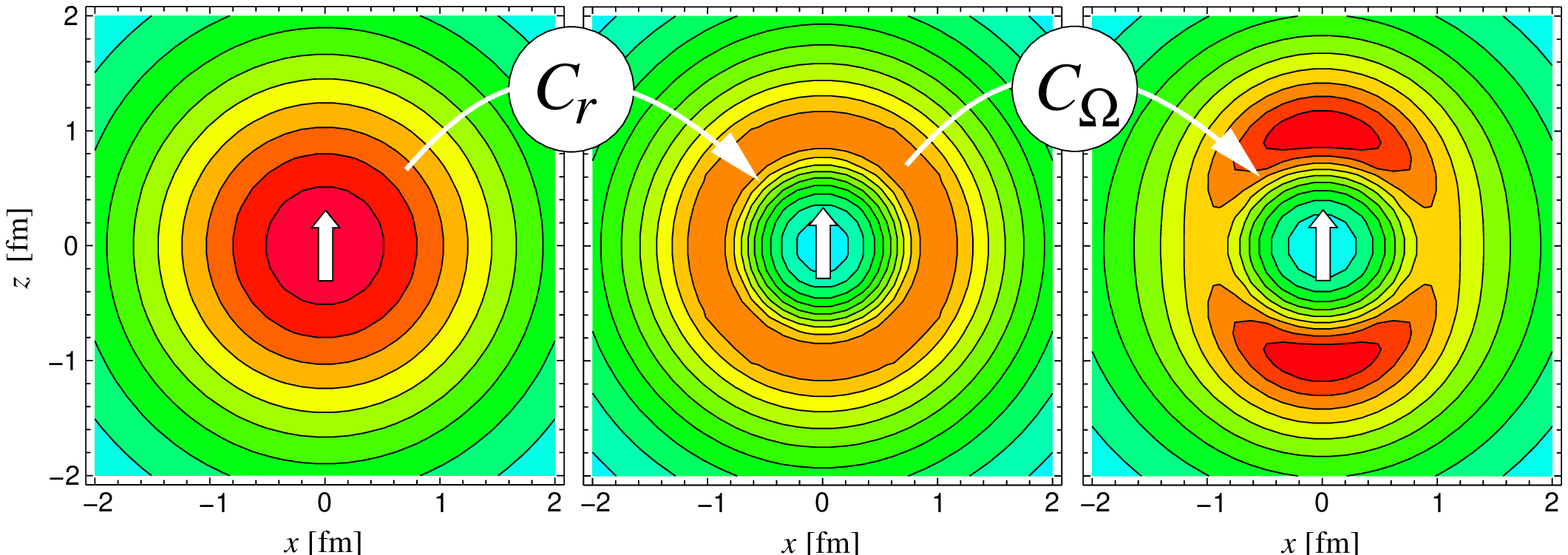}\\[5mm]
    \includegraphics[angle=0,width=0.63\linewidth]{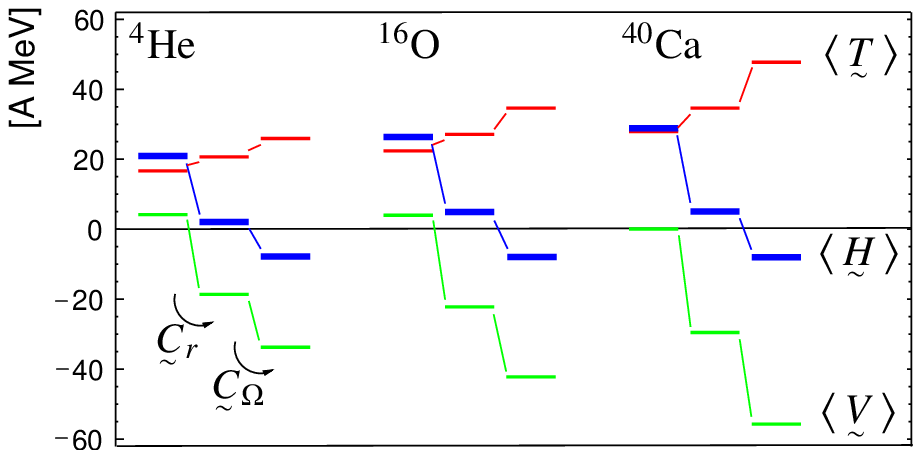}
  \end{center}
  \caption{Upper part: two-body density
    $\rho^{(2)}_{S,T}(\vec{r}_1\!\!-\!\vec{r}_2)$ of
    $\chemical{4}{He}$ for a pair of nucleons with isospin $T\sheq0$
    and parallel spins, $S\sheq1, M_S\sheq1$. Arrow indicates spin
    direction and $(x,y,z)=(\vec{r}_1\!\!-\!\vec{r}_2)$ relative
    distance vector.  Lower part: corresponding kinetic, potential and
    total energies per particle of $\chemical{4}{He}$,
    $\chemical{16}{O}$ and $\chemical{40}{Ca}$, without, with radial,
    and with radial and tensor correlations (Bonn-A).}
\label{fig:BonnA-Energies}
\end{figure}

When applied to a many-body state $\ket{\Psi}$ the radial correlator
$\op{C}_r$ shifts all particle pairs $(i,j)$ radially away from each
other whenever they are too close, i.e. inside the range of the
repulsive core.  The strength function $s(r_{\idxij})$ controls the
amount of the radial shift and is optimized to the potential under
consideration. $\op{p}_{r\idxij}$ is the radial component of the
relative momentum. The effect of the transformation
$\ket{\Psi}\rightarrow \op{C}_r\ket{\Psi}$ is shown in the upper part
of Fig.~\ref{fig:BonnA-Energies} where the two-body density
$\rho^{(2)}_{S,T}$ is displayed as a function of the distance vector
$(\vec{r}_1\!-\!\vec{r}_2)$ between two nucleons in
$\chemical{4}{He}$. The $\rho^{(2)}_{S,T}$ on the l.h.s. is calculated
with the shell-model state $\ket{(1s_{1/2})^4}$ that is just a product
of 4 Gaussians.  It has a maximum at zero distance which is in
contradiction to the short ranged repulsion of the interaction.  This
inconsistency is removed by the action of the radial correlator
$\op{C}_r$ that moves density out of the region where the potential is
repulsive.  The corresponding kinetic, potential and total energies
are displayed in the lower part of the figure for three nuclei.  The
radially correlated kinetic energy $\expect{\op{C}_r^\dagger \op{T}
  \op{C}_r}$ increases somewhat compared to $\expect{\op{T}}$ but this
is overcompensated by the gain of about $-25\:\MeV$ per particle in
the correlated potential energy.  Nevertheless the nuclei are still
unbound.

The tensor correlations are induced by $\op{C}_\Omega$ where the
tensor operator in the exponent (Eq. (\ref{eq:opcom})) creates shifts
perpendicular to $\vec{r}_{\idxij}$.  The amount of the displacement
depends on the spin directions $\op{\vec{\sigma}}_i$ and
$\op{\vec{\sigma}}_j$ of the particles relative to their distance
vector.  The operator $\op{\vec{p}}_{\Omega \idxij}= \op{\vec{p}}_{\idxij} -
\op{\vec{p}}_{r \idxij}$, called orbital relative momentum, is the
component of $\op{\vec{p}}_{\idxij}$ perpendicular to
$\op{\vec{r}}_{\idxij}$.  The overall strength of the tensor correlations
is controlled by $\vartheta(r_{\!\idxij})$ and allows, for example, to
map a purely $l\!=\!0$ deuteron wave function onto the exact one which
includes an $l\!=\!2$ component and thus all tensor correlations [1].
The application of the tensor correlator $\op{C}_\Omega$ leads to the
two-body density depicted in the right hand contour plot of
Fig.~\ref{fig:BonnA-Energies}.  One may visualize the action of
$\op{C}_\Omega$ as a displacement of probability density from the
'equator' to both 'poles', where the spin of the $S\sheq 1$ component
of the nucleon pair defines the 'south-north' direction.  Again this
costs kinetic energy but now the many-body state is in accord with the
tensor interaction and one gains the binding needed to end up with
about -8~MeV per particle, see lower part of
Fig.~\ref{fig:BonnA-Energies}.

\section{Choice of correlation functions}

Like the interaction the correlators are decomposed into the four spin
isospin channels $S\sheq0,1;\ T\sheq0,1$ and the corresponding
strength functions $s(r)$ and $\vartheta(r)$ are adjusted separately.
As the repulsion is of short range the optimal radial shift functions
can be found by minimizing the energy with respect to $s(r)$ in the
corresponding channel.  The result depends only weakly on the choice
of the system, a constant trial state in the two-body system gives
practically the same $s(r)$ as minimizing the energy of the deuteron
or $\chemical{4}{He}$.
\begin{figure}[b]
  \begin{center}
    \includegraphics[angle=0,width=0.5\linewidth]{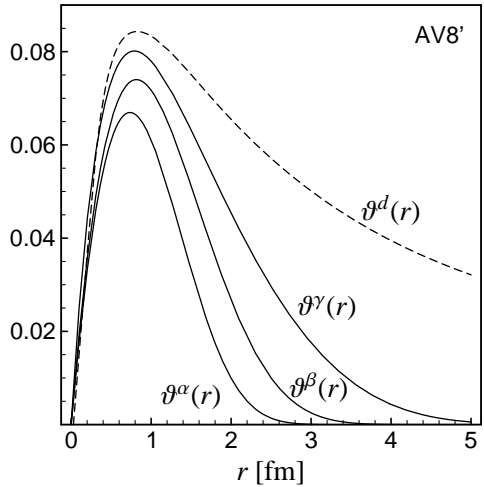}
  \end{center}
  \caption{Strength of tensor correlation as function of particle distance.}
  \label{fig:theta}
\end{figure}

Unlike the repulsive core the tensor part of the interaction, that is
due to pion exchange, is of long range.  Therefore, adjusting
$\vartheta(r)$ by mapping a pure $l\sheq0$ trial state for the
deuteron to the exact eigenstate leads to the very long ranged tensor
correlator strength $\vartheta^d(r)$ shown in Fig.~\ref{fig:theta}.
When used in systems with more than two particles this correlator is
not useful as it induces large three- and more-body parts in the
correlated interaction ($\cop{H}^{[3]},\ldots$) that are very
complicated to handle. Therefore we restrict the range of the tensor
correlations and consider in the following the three $\vartheta(r)$
labeled by $\alpha$, $\beta$ and $\gamma$ shown in
Fig.~\ref{fig:theta}.  With those we take care of the short range part
of the tensor correlations. The long range part must then be described
by the many-body state $\ket{\Psi}$. The presence of the other
particles will destroy at least partially the alignment of spins of a
particle pair at larger distances (spin frustration) so that
$\ket{\Psi}$ can be again a rather simple many-body state.

\section{Correlated interaction in momentum space}

\begin{figure}[b]
 \includegraphics[width=1.0\columnwidth]{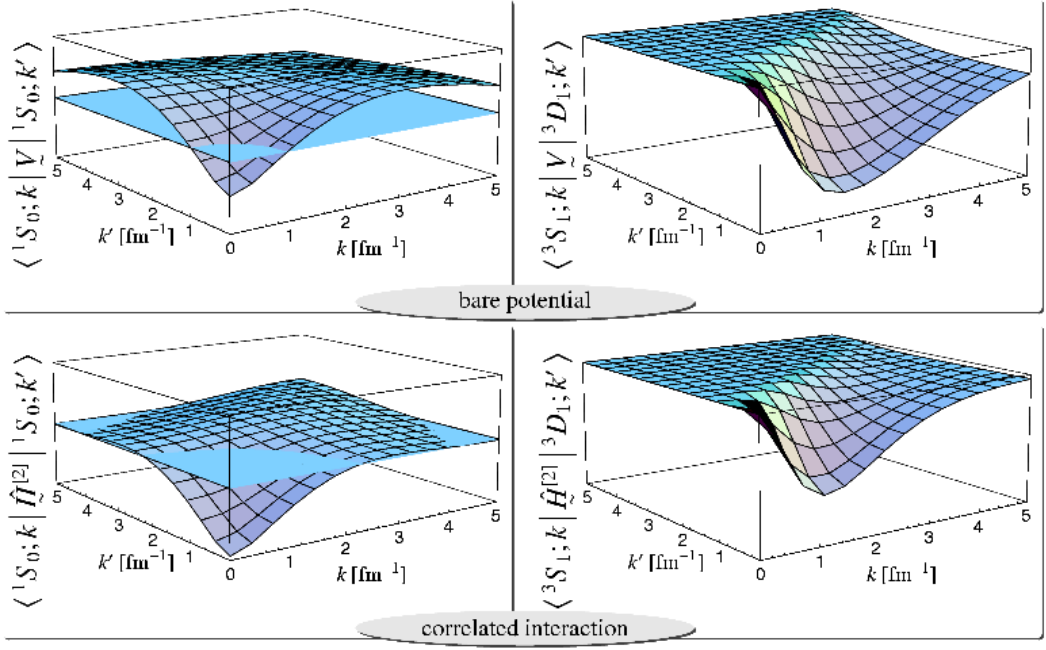}
 \caption{Matrix elements in momentum representation for the
  uncorrelated and correlated Argonne V8' potential.  Left: $^1S_0$
  channel, plane denotes zero, vertical range from -2 to 2 $\fm^{-1}$.
  Right: tensor mixing between $^3S_1$ and $^3D_1$ channel, vertical
  range from -1.5 to 0 $\fm^{-1}$.}
\label{fig:Vkk}
\end{figure}

In Fig.~\ref{fig:Vkk} we display in momentum representation the matrix
elements of the bare Argonne V8' potential with those of the
corresponding correlated interaction $\coptwo{H}$. The left column is
for the $^1S_0$ channel where due to $S\sheq0$ only the radial
correlator acts.

It is obvious that the goal of pre-diagonalization is achieved.
Beyond momentum transfers of about 2~fm$^{-1}$ the off-diagonal matrix
elements calculated with correlated states are close to zero.  Our
result is in good agreement with $V_{low-k}$ obtained with
renormalization group methods [1,4].  In the right column the matrix
elements between the $l\sheq 0$ and $l\sheq2$ triplet channels are
shown. Here only the tensor components of the interaction contribute.
For the correlated state we use the correlation strength
$\vartheta^\alpha(r)$ shown in Fig.~\ref{fig:theta}. Despite the
restricted range the correlator achieves a substantial reduction of
the matrix elements. With the long ranged correlator $\vartheta^d(r)$
derived from the deuteron the off-diagonal matrix elements vanish
completely.

\section{No-core shell model calculations}

\begin{figure}[b]
  \begin{center}
    \includegraphics[angle=0,width=0.9\linewidth]{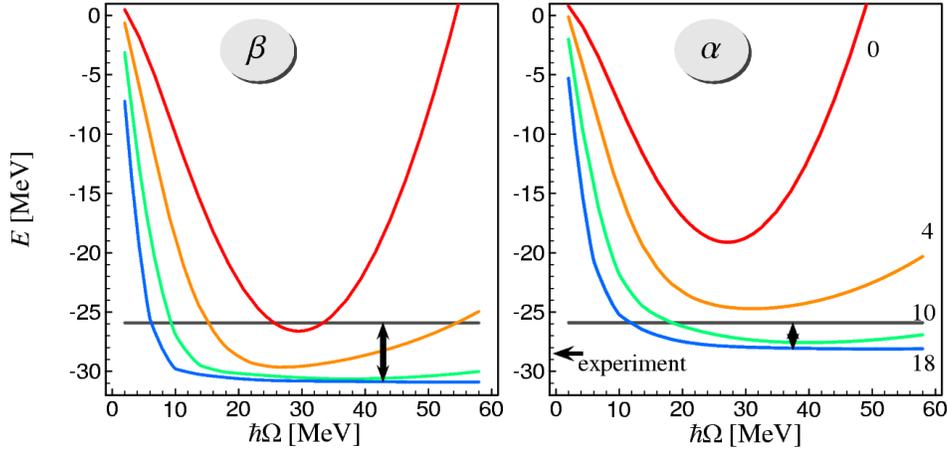}
  \end{center}
  \caption{Ground state energy of $\chemical{4}{He}$ calculated with
    Argonne V8' as function of harmonic oscillator parameter $\hbar
    \Omega$ and number of major shells included.  L.h.s.: Hamiltonian
    correlated with tensor correlator of range $\beta$, r.h.s.:
    shorter range $\alpha$. Horizontal line: benchmark calculations
    [5].}
  \label{fig:nocorehe4}
\end{figure}

In order to test the UCOM we use the correlated Hamiltonian in
two-body approximation $\coptwo{H}$ with the no-core shell model code
of P.  Navr{\'a}til [6] and compare with benchmark calculations for
$\chemical{4}{He}$ [5].  

In Fig.~\ref{fig:nocorehe4} we show the ground state energy as a
function of the oscillator parameter for $0$, $4$, $10$ and $18\:\hbar
\Omega$ excitations calculated with tensor correlators of range
$\beta$ and $\alpha$. The correlated Hamiltonian obtained with the
medium ranged correlator $\beta$ gives with a single Slater
determinant ($0\:\hbar \Omega$) a minimum in energy at almost the
exact energy (horizontal line).  Admixing more and more configurations
lowers the energy and convergence is reached at about $8 \hbar
\Omega$.  One should keep in mind that the bare interaction needs
many-body spaces that are orders of magnitude larger. For example at
$\hbar\Omega = 25\:\MeV$ the binding energy calculated with
excitations up to $18\:\hbar\Omega$ included is only $-6\:\MeV$.  If
we use the short-ranged tensor correlator $\alpha$ the optimal single
Slater determinant is $7\: \MeV$ above the exact eigenvalue.  As can
be seen in Fig.~\ref{fig:nocorehe4} convergence is not as fast but the
overbinding is reduced to about $2\:\MeV$. This difference to the
reference calculations is attributed to the missing contributions from
$\cop{H}^{[3]}$ and $\cop{H}^{[4]}$.

\section{Mean-field calculations in FMD basis}

\begin{figure}[b]
  \begin{center}
   \includegraphics[angle=0,width=0.31\textwidth]{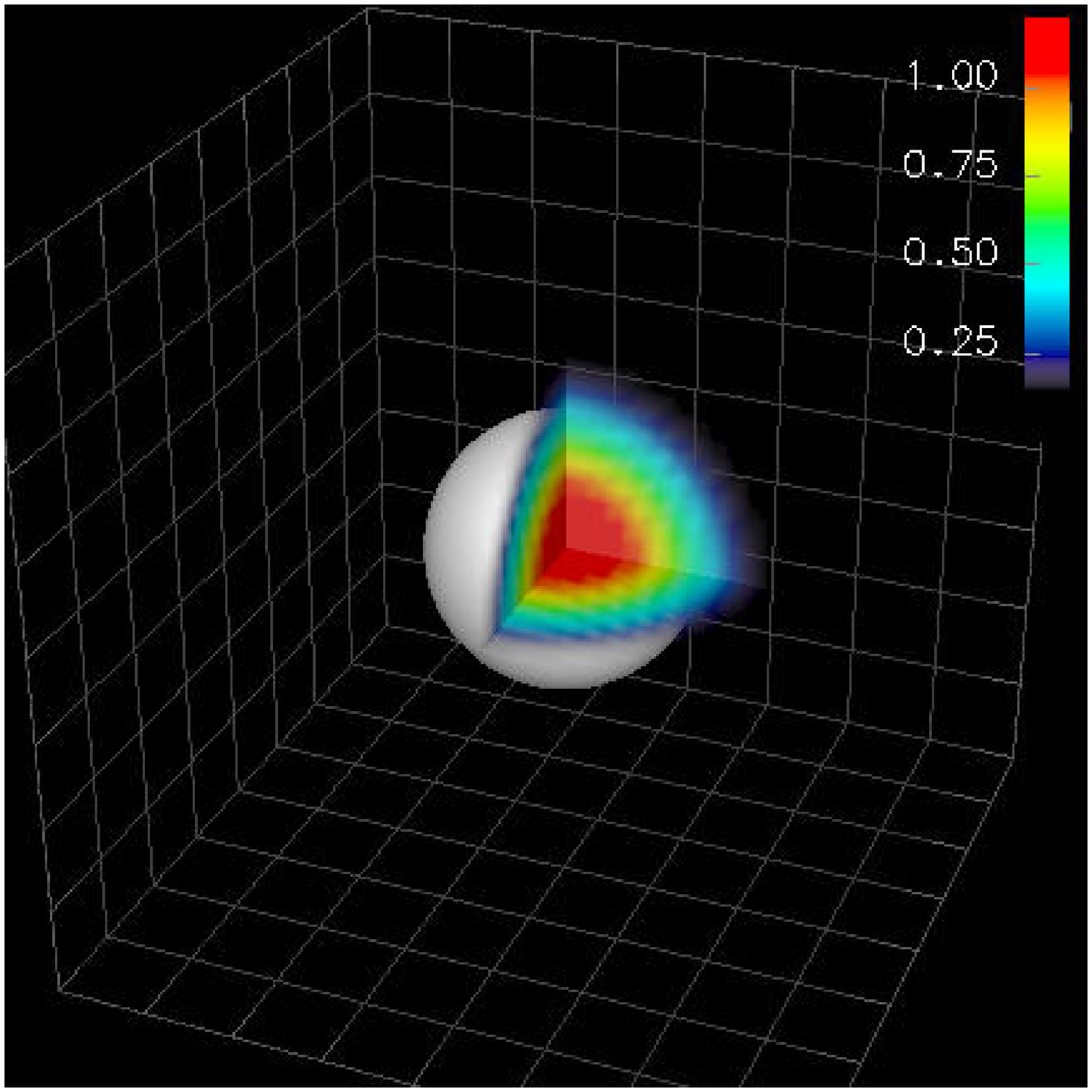}
   \includegraphics[angle=0,width=0.31\textwidth]{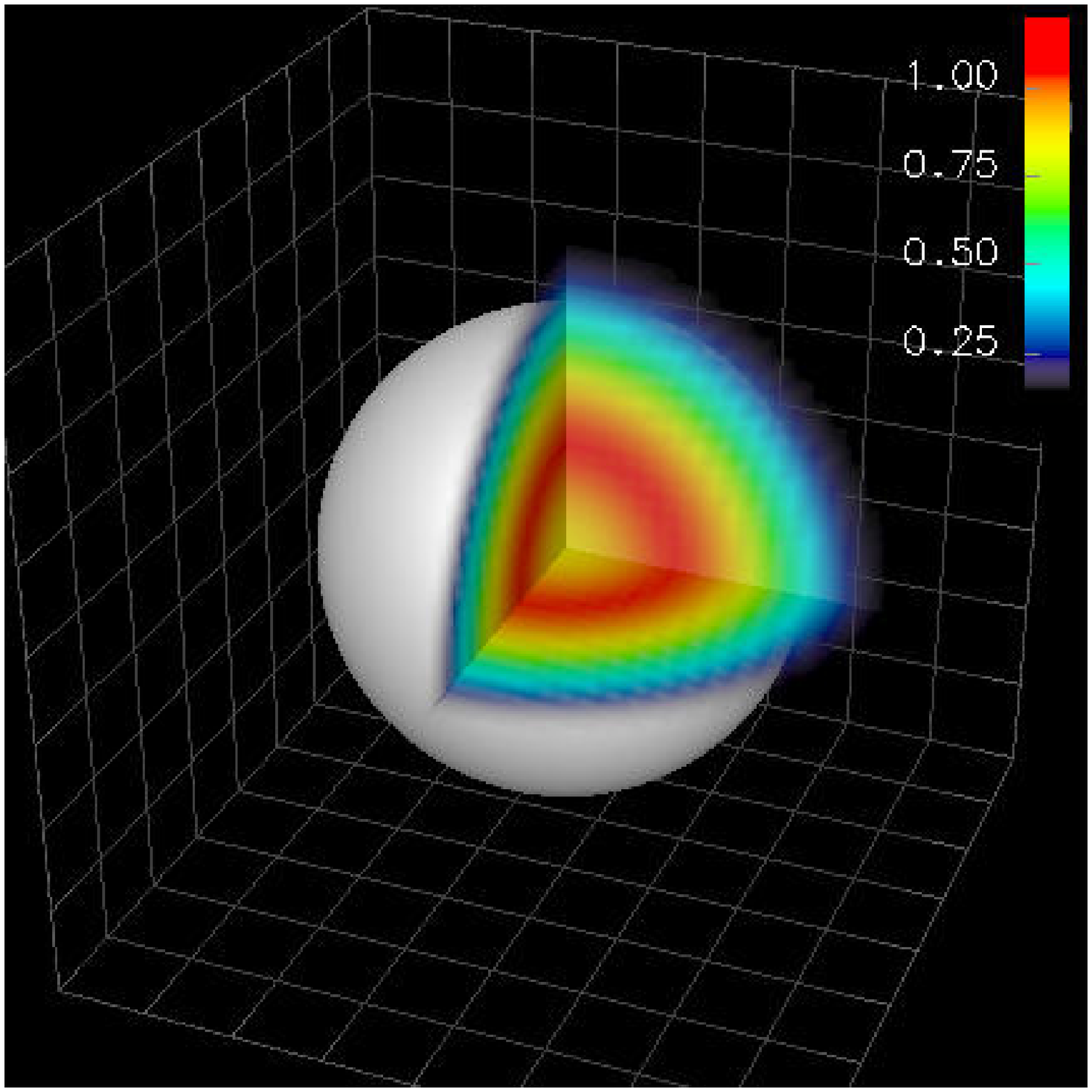}
   \includegraphics[angle=0,width=0.31\textwidth]{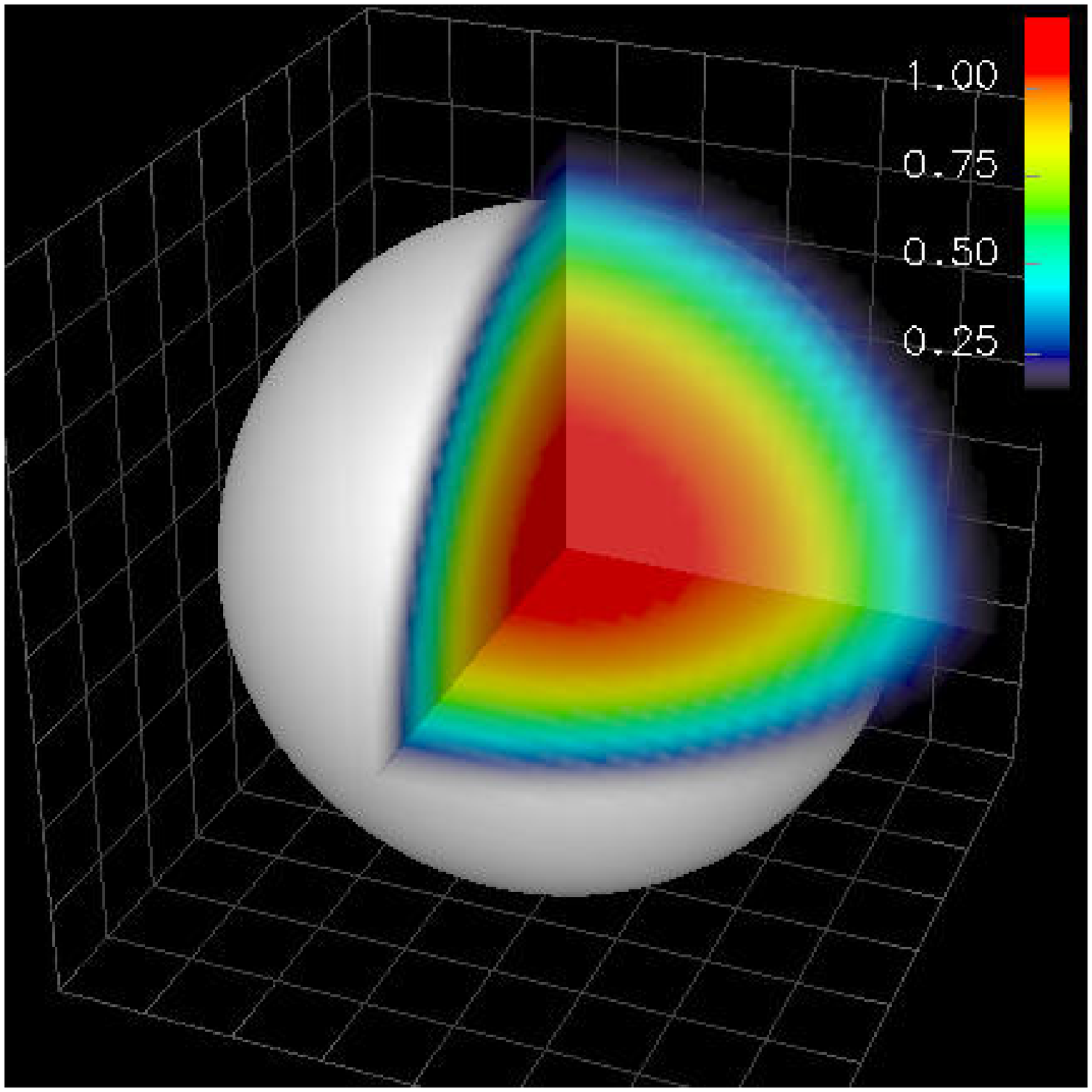}\\[1mm]
   \includegraphics[angle=0,width=0.31\textwidth]{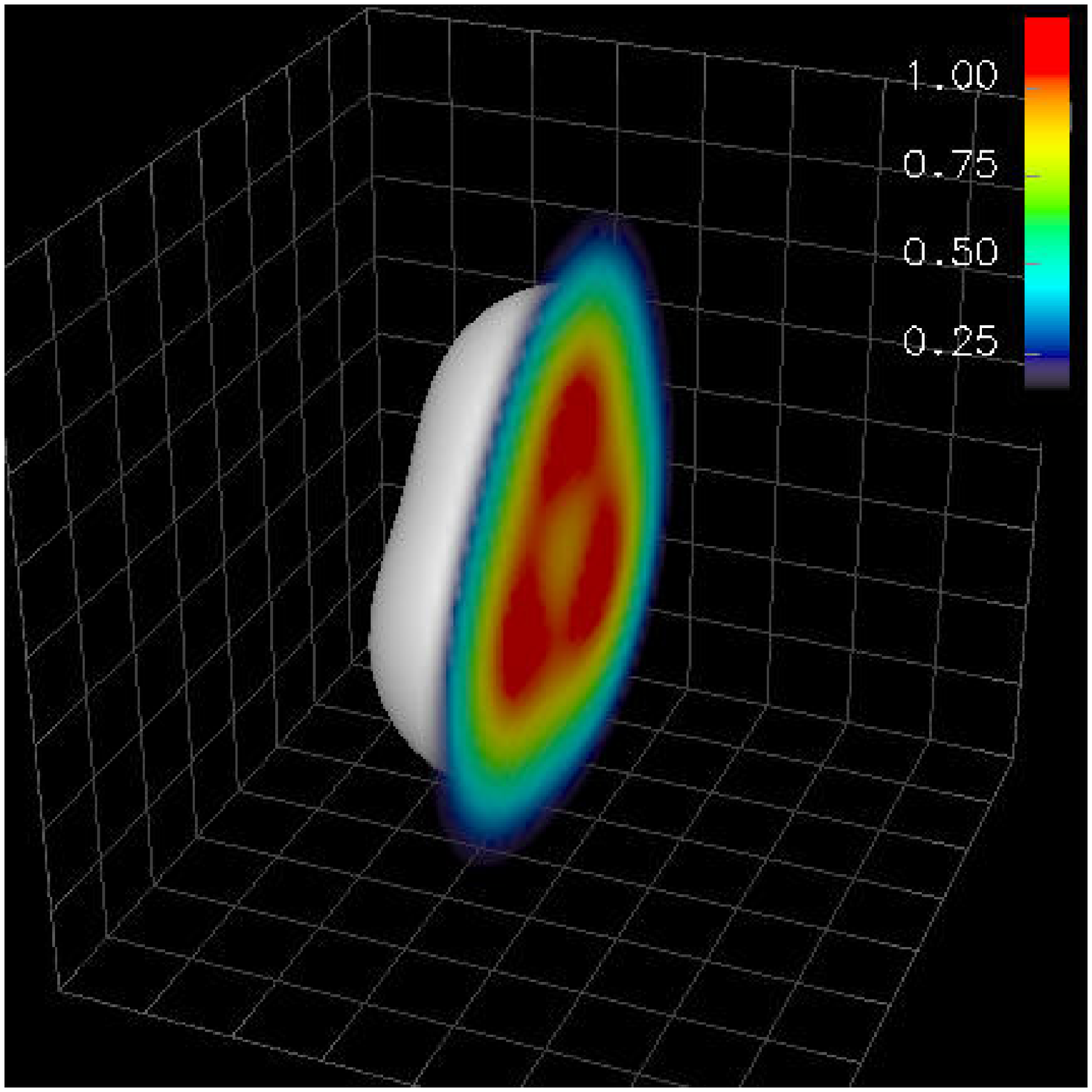}
   \includegraphics[angle=0,width=0.31\textwidth]{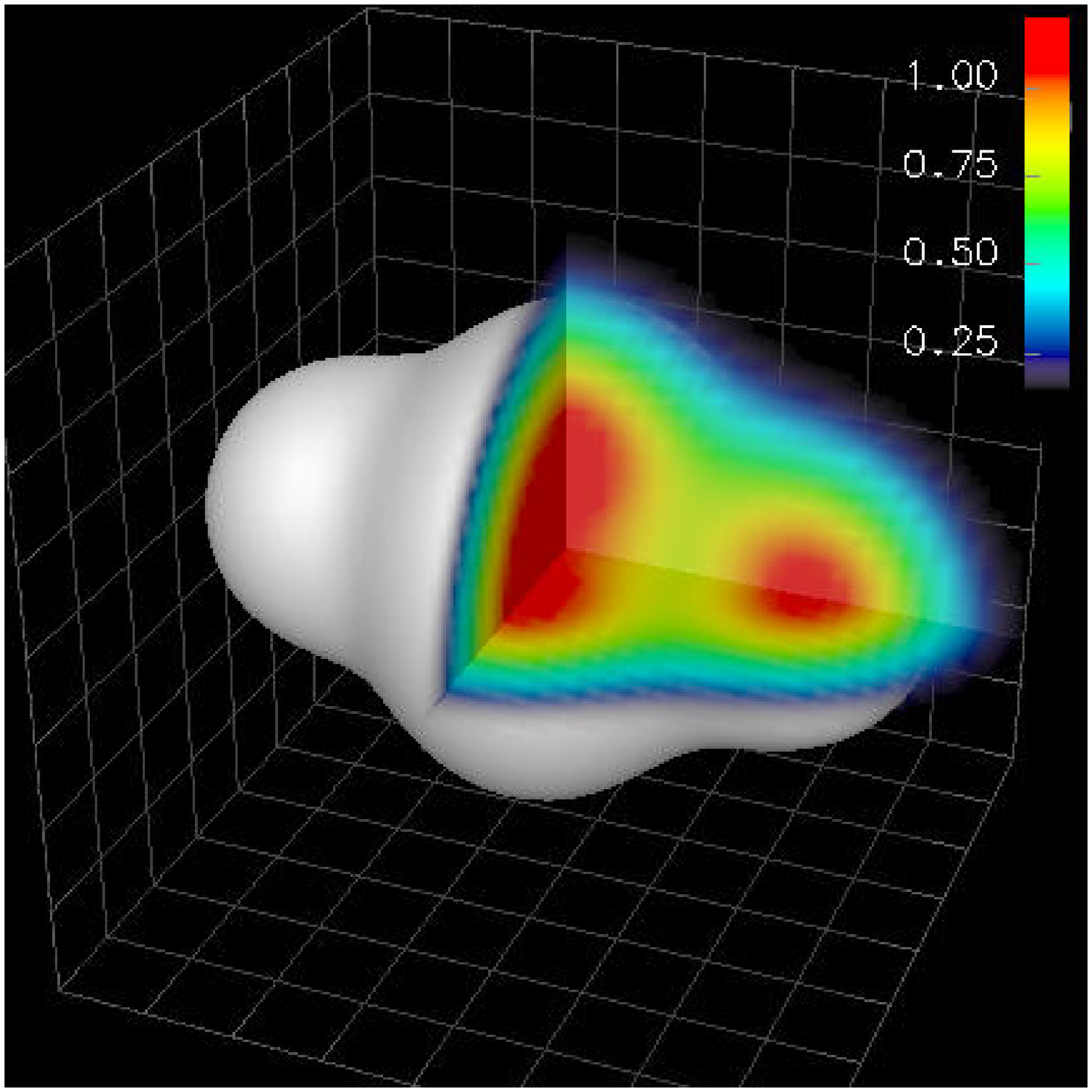}
   \includegraphics[angle=0,width=0.31\textwidth]{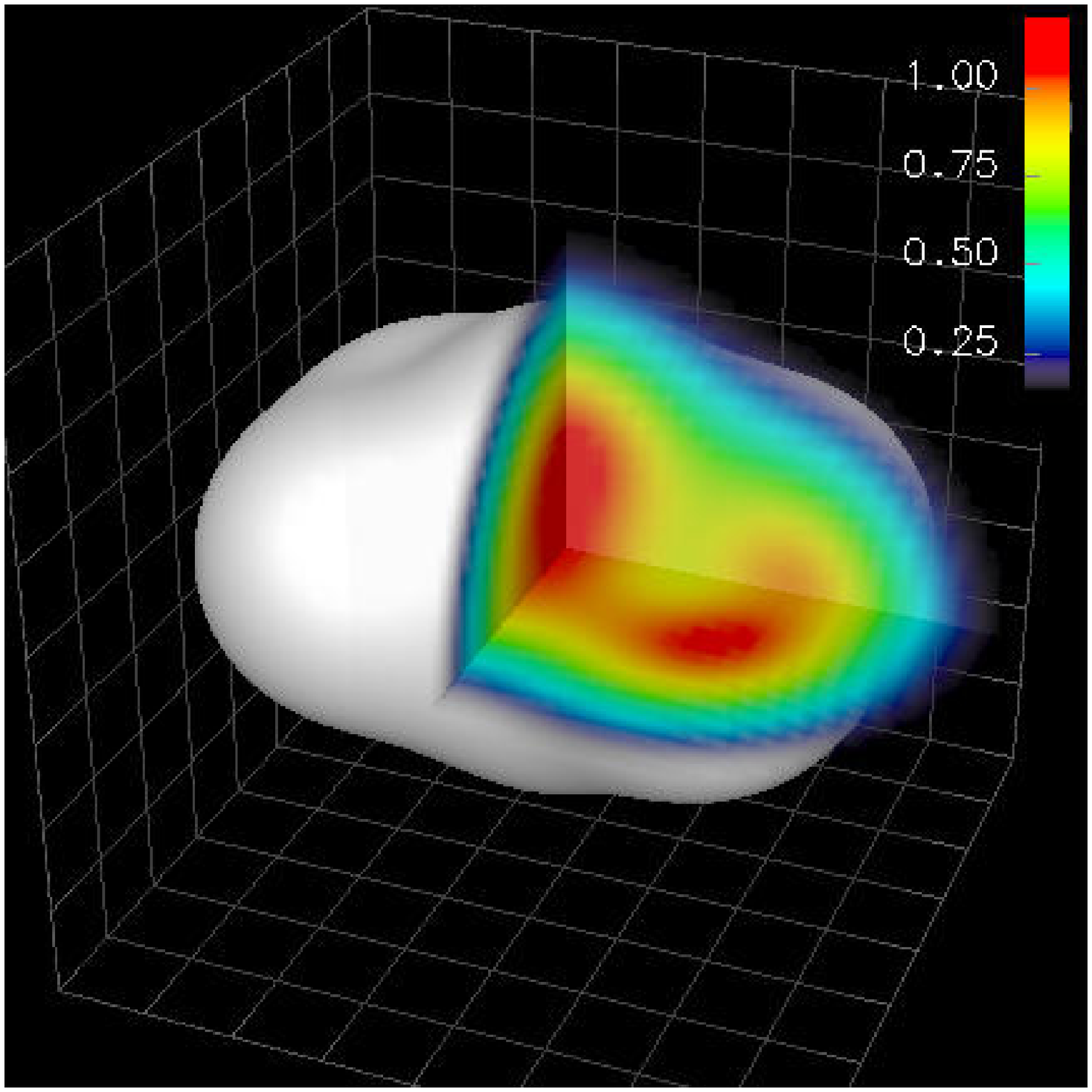}
 \end{center}
 \caption{One-body densities of $\chemical{4}{He}$,
   $\chemical{16}{O}$, $\chemical{40}{Ca}$ and $\chemical{12}{C}$,
   $\chemical{20}{Ne}$, $\chemical{24}{Mg}$.}
 \label{fig:SelectedNuclei}
\end{figure}

For heavier nuclei no-core shell model calculations are not feasible
and we perform mean-field calculations in the framework of the
Fermionic Molecular Dynamics (FMD) model [3]. The trial state in this
approach consists of a single Slater determinant with single-particle
states that are parametrized as a single or a sum of two Gaussians.
As Hamiltonian we use the correlated Bonn-A interaction that is
complemented by a momentum-dependent two-body correction term. This
correction term has to simulate the effect of the missing three- and
more-body terms of the correlated interaction and the effect of the
genuine three-body forces. It is adjusted to reproduce the binding
energies and radii of the doubly-magic nuclei $\chemical{4}{He}$,
$\chemical{16}{O}$ and $\chemical{40}{Ca}$.  Minimizing the binding
energy with respect to the parameters of the single-particle states
yields intrinsic states. The resulting one-body densities are shown
for some nuclei in Fig.~\ref{fig:SelectedNuclei}.

\begin{figure}[t]
  \begin{center}
    \includegraphics[angle=0,width=0.97\textwidth]{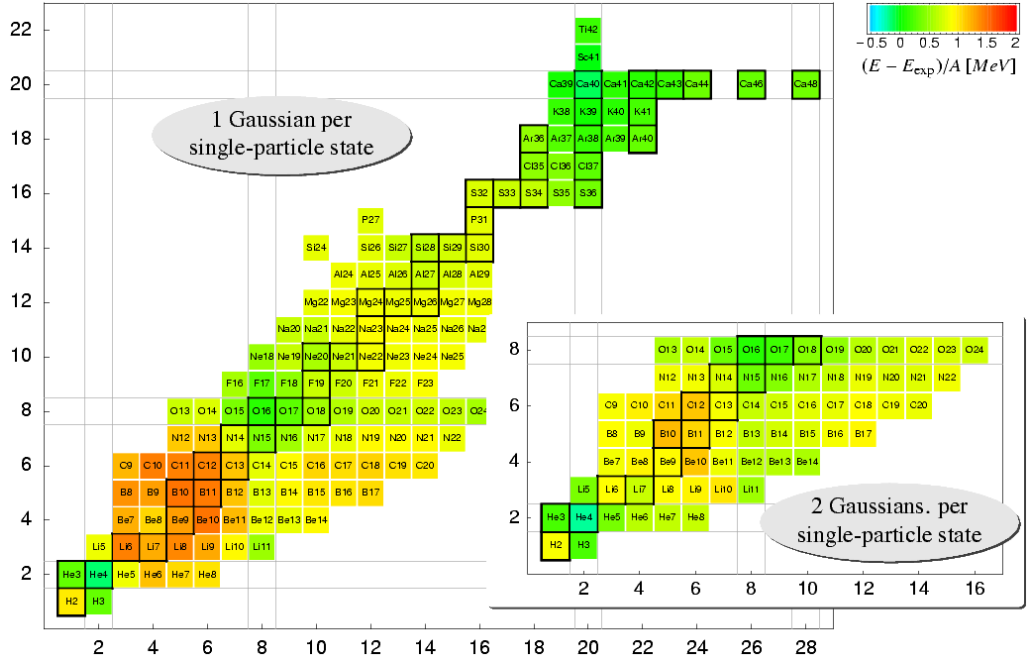}
  \end{center}
  \caption{Deviation of intrinsic energies from experimental ground state
    energies. Insert: two Gaussians per single particle state.}
  \label{fig:FMDnuclchart}
\end{figure}

Besides the spherical nuclei
$\chemical{4}{He}$, $\chemical{16}{O}$ and $\chemical{40}{Ca}$ we
obtain also an axial $\chemical{20}{Ne}$ and a triaxial
$\chemical{24}{Mg}$. The energies of the intrinsic states are shown in
Fig.~\ref{fig:FMDnuclchart}. For the $p$-shell we achieve a
substantial improvement by using single-particle states with two
Gaussians per nucleon. For the doubly-magic nuclei the spherical
intrinsic states can be regarded as a good approximation of the ground
states. On the other hand nuclei between closed shells are
intrinsically deformed so that a projection on good angular momentum
should be performed before comparing with experimental binding energies.

Configuration mixing calculations taking into account rotations and
vibrations are under way and should allow a description of the low
lying spectra. First results are promising.

\vspace{2mm}

We are grateful to Petr Navr{\'a}til for providing his no-core shell
model code used in the $\chemical{4}{He}$ calculations.

\vspace{2mm}

\noindent
{[1]} T. Neff, H. Feldmeier, NP{\bf A713} (2003) 311 \\
{[2]} H. Feldmeier et al., NP{\bf A632} (1998) 61 \\
{[3]} H. Feldmeier, J. Schnack, Rev.Mod.Phys. {\bf 72} (2000) 655 \\
{[4]} S. Bogner et al., nucl-th/0108041\\
{[5]} H. Kamada, et al., PR{\bf C64} (2001) 044001\\
{[6]} P. Navr{\'a}til et al., PR{\bf C61} (2000) 61044001

\end{document}